Stabilization of high-$T_c$ phase of BiS$_2$-based superconductor LaO$_{0.5}$F$_{0.5}$BiS$_2$ using high-pressure synthesis

Yoshikazu Mizuguchi[1*], Takafumi Hiroi[1], Joe Kajitani[1], Hiroshi Takatsu[2], Hiroaki Kadowaki[2] and Osuke Miura[1]

*1. Department of Electrical and Electronic Engineering, Tokyo Metropolitan University, 1-1, Minami-osawa, Hachioji, Japan*
*2. Department of Physics, Tokyo Metropolitan University, 1-1, Minami-osawa, Hachioji, Japan*

*Corresponding author: Yoshikazu Mizuguchi (mizugu@tmu.ac.jp)

Abstract

High-quality polycrystalline samples of LaO$_{0.5}$F$_{0.5}$BiS$_2$ were obtained using high-pressure synthesis technique. The LaO$_{0.5}$F$_{0.5}$BiS$_2$ sample prepared by heating at 700 ºC under 2 GPa showed superconductivity with superconducting transition temperatures ($T_c$) of $T_c^{\mathrm{onset}}$ = 11.1 and $T_c^{\mathrm{zero}}$ = 8.5 K in the electrical resistivity measurements and $T_c^{\mathrm{onset}}$ = 11.5 and $T_c^{\mathrm{irr}}$ = 9.4 K in the magnetic susceptibility measurements, which are obviously higher than those of the LaO$_{0.5}$F$_{0.5}$BiS$_2$ polycrystalline samples obtained using conventional solid-state reaction. It was found that the high-$T_c$ phase can be stabilized under high pressure and relatively-low annealing temperature. X-ray diffraction analysis revealed that the high-$T_c$ phase possessed a small ratio of lattice constants of $a$ and $c$, $c/a$.

The $BiS_2$-based superconductors are novel layered superconductors [1,2]. The crystal structure is basically composed of an alternate stacking of double $BiS_2$ conduction layers and blocking layers such as $RE_2O_2$ (RE: Rare earth), $Sr_2F_2$ or $Bi_4O_4(SO_4)_{1-x}$ layers[1-11]. Because of the layered structure similar to the cuprtes and the Fe-based family [12,13], appearance of novel mechanisms of superconductivity and high superconducting transition temperature ($T_c$) has been expected. Band calculations suggested that the parent material is a band insulator, and the $BiS_2$-based materials could be superconductive when electron carriers were generated at the Bi-$6p$ orbitals within the $BiS_2$ conduction layers [14].

$LaO_{1-x}F_xBiS_2$ is a typical $BiS_2$-based superconductor [2]. The parent phase $LaOBiS_2$ is semiconducting, and partial substitution of O by F generates electron carriers within the $BiS_2$ layers. The superconducting transition temperature $T_c$ of $LaO_{1-x}F_xBiS_2$ becomes highest at $x \sim 0.5$. The $T_c$ of the polycrystalline samples prepared by a conventional solid-state reaction (annealing in an evacuated quartz tube) is less than 3 K. However, the $T_c$ can be largely enhanced up to 10 K by applying high pressure (HP) using a HP cell [15-17]. Recently, Tomita et al. showed the possibility of structural transition from tetragonal (low-$T_c$) to monoclinic (high-$T_c$) on the basis of the results of X-ray analysis under HP and the molecular dynamics simulations of density functional theory calculations. Another way to enhance the $T_c$ in $LaO_{1-x}F_xBiS_2$ is HP annealing. The sample annealed at 600 ºC under 2 GPa showed $T_c^{onset}$ = 10.6 and $T_c^{zero}$ = 7.8 K [2,18]. The enhancement of $T_c$ was explained by the uniaxial lattice strain along the $c$ axis [19]. These scenarios of the appearance of high-$T_c$ phase in $LaO_{1-x}F_xBiS_2$ concluded in refs. 17 and 19 are basically consistent. However, to clarify the correlation between superconductivity and the crystal structure and to investigate physical properties of the high-$T_c$ phase, the preparation of homogeneous samples with $T_c$ higher than 10 K is much important. Here we report the HP synthesis of high-quality $LaO_{0.5}F_{0.5}BiS_2$ polycrystalline samples with $T_c^{onset}$ = 11.1 and $T_c^{zero}$ = 8.5 K in the electrical resistivity measurements and $T_c^{onset}$ = 11.5 and $T_c^{irr}$ = 9.4 K in the magnetic susceptibility measurements.

Polycrystalline samples of $LaO_{0.5}F_{0.5}BiS_2$ were prepared using a HP synthesis technique with a cubic-anvil-type 180 ton press. The starting materials are $La_2S_3$ (99.9 %), $Bi_2O_3$ (99.99 %), $BiF_3$ (99.9 %) and $Bi_2S_3$ powders and Bi (99.99 %) grains. The $Bi_2S_3$ powder was prepared by reacting Bi and S (99.99 %) grains. The mixture with a nominal composition of $LaO_{0.5}F_{0.5}BiS_2$ was well-mixed and pressed into a pellet. The sample was sealed into a BN sample space and inserted into a HP cell. The HP cell was pressed with a pressure of 2 GPa and heated with various annealing temperatures and times. The detailed annealing conditions are listed in Table 1. The temperature dependence of electrical resistivity below 300 K was measured using the four-terminal method. The DC magnetic susceptibility was measured using a superconducting quantum interference device (SQUID) magnetometer after both zero-field cooling (ZFC) and field cooling (FC). The $T_c^{onset}$ in the susceptibility measurement was estimated to be a temperature where the susceptibility began to decrease. The error of the $T_c^{onset}$ in the phase diagram (Fig. 4) was estimated by considering the real value of the susceptibility error at normal conducting states around 15 K. For $T_c^{irr}$, the reliable estimation of error was quite difficult due to drastic change of susceptibility around $T_c^{irr}$. Hence we did not plot in the phase diagram. The powder X-ray powder diffraction (XRD) experiments were carried out using a Rigaku SmartLab powder diffractometer equipped with a CuK$_{\alpha 1}$ monochrometer. To compare XRD patterns of various samples precisely, the standard Si powder was mixed with the samples, and used as the reference of the calibration of the peak shift.

The highest $T_c$ and shielding property were obtained for the LaO$_{0.5}$F$_{0.5}$BiS$_2$ sample prepared with heating at 700 ºC for 1 h under 2 GPa. The temperature dependence of electrical resistivity for the HP sample was displayed in Fig. 1(a). For comparison, resistivity data of the solid-state-reacted LaO$_{0.5}$F$_{0.5}$BiS$_2$ polycrystalline sample was plotted together. The $T_c$ of the HP sample is obviously higher than that of the solid-state-reacted sample. The $T_c^{onset}$ and $T_c^{zero}$ are 11.1 and 8.5 K in the electrical resistivity measurement. The onset temperature was defined as a temperature where the resistivity began to decrease due to the evolution of the superconducting states as shown in Fig. 1(c). Figure 1(b) shows the temperature dependence of magnetic susceptibility for the HP and solid-state-reacted samples. In the susceptibility measurements, $T_c^{onset}$ and $T_c^{irr}$ are 11.5 and 9.4 K, respectively. The $T_c^{irr}$ is a bifurcation temperature between the ZFC and FC curves as shown in Fig. 1(d).

Figure 2 shows the XRD pattern of the HP sample prepared with heating at 700 ºC for 1 h under 2 GPa (HP-700 ºC). The Miller indices are displayed over the corresponding peaks. The obtained XRD pattern shows an almost single phase, except for the tiny impurity phase of Bi$_2$S$_3$. By converting peak positions of the (200) and (003) reflections to the lattice constants $a$ and $c$ for the tetragonal unit cell, we obtained $a$ = 4.0748(1) Å and $c$ = 13.2830(10) Å for the HP-700 ºC sample. As summarized in Table 1, the single-phase samples were obtained when the heating temperature was above 700 ºC. This fact indicates that the formation of the LaOBiS$_2$-type structure requires a heating temperature higher than 650 ºC under a high pressure of 2 GPa. For the HP sample prepared at 800 ºC (HP-800 ºC), single phase was obtained by heating only for 0.5 h. However the $T_c$ of the HP-800 ºC sample becomes lower than those of the HP-700 ºC sample.

In Fig. 3, we plot XRD profiles around the (003) and (200) peaks for the solid-state-reacted sample, HP-700 ºC sample (the high-$T_c$ phase) and HP-800 ºC sample (the low-$T_c$ phase), respectively. As in the case of the HP-700 ºC sample, the lattice constants of the HP-800 ºC sample are obtained as $a$ = 4.0727(1) Å and $c$ = 13.2920(10) Å. The lattice constants of the solid-state-reacted sample are $a$ = 4.0709(1) Å and $c$ = 13.3475(3) Å. One can see the broadening of the peaks when the $T_c$ of the sample is increased. Moreover, the peak positions of the (003) and (200) reflections monotonically increases and decreases, respectively. Therefore, we think that there is a close correlation between the lattice parameters and the appearance of the high-$T_c$ phase in the BiS$_2$-based superconductor. We discuss its detail below.

Firstly, we compare the solid-state-reacted sample and the HP-700 ºC sample. The position of the (003) peak for the HP-700 ºC sample is higher than that of the solid-state-reacted sample. This indicates that the length of the $c$ axis for the HP-700 ºC sample is clearly shorter than that of the solid-state-reacted sample. In contrast, peak position of the (200) peak for the HP-700 ºC sample is lower than that of the solid-state-reacted sample. This indicates that the length of the $a$ axis in the HP-700 ºC sample is longer than that of the solid-state-reacted sample. Namely, a small $c/a$ ration is important for the appearance of the high-$T_c$ phase. These tendencies are basically consistent with the uniaxial contraction along the $c$ axis observed in the HP-post-annealing studies [2,18,19]. Although the previous X-ray studies for the solid-state-reacted LaO$_{0.5}$F$_{0.5}$BiS$_2$ sample under HP suggested that the monoclinic phase appeared under HP, we did not observe a clear evidence of the structural phase transition to the monoclinic structure in the present studies on the HP-synthesized samples. Nevertheless, it should be noted that the (003) and (200) peaks are broader than those of the solid-state-reacted sample, implying a little but finite lattice distortion into the HP-700 ºC sample. Therefore, we speculate that the high-$T_c$ phase of the LaO$_{0.5}$F$_{0.5}$BiS$_2$ requires an intermediate lattice structure between tetragonal and monoclinic structure.

Next, we discuss the differences in the superconducting properties and crystal structure between the HP-700 °C sample and HP-800 °C sample. As shown in Fig. 1, the $T_c$ of the HP-700 °C sample is the highest, $T_c^{onset}$ = 11.5 K and $T_c^{irr}$ = 9.4 K, and the shielding volume fraction estimated using a value of ZFC magnetization at 2 K is almost 100 %. In contrast, the HP-800 °C sample does not show bulk superconductivity; the estimated $T_c$ are $T_c^{onset}$ = 8.6 K and $T_c^{irr}$ = 4.1 K and the shielding volume fraction calculated using a value of ZFC magnetization at 2 K is ~3.5 %. As shown in Fig. 3(a), the position of the (003) peak for the HP-800 °C sample is slightly lower than that of the HP-700 °C sample. In contrast, the position of the (200) peak is higher than that of the HP-700 °C sample as shown in Fig. 3(b). This means that the HP-800 °C sample possesses a shorter $a$ axis and a longer $c$ axis than those of the HP-700 °C sample. To visualize the relationship between $T_c$ and the $c/a$ ratio, the data for the solid-state-reacted, the HP-700 °C and the HP-800 °C samples are plotted in Fig. 4. It is clear that the appearance of higher-$T_c$ superconductivity requires the smaller $c/a$ ratio. Furthermore, the peaks of the HP-800 °C sample are sharper than those of the HP-700 °C sample. Thus, the HP-800 °C sample could be considered as almost a normal tetragonal phase. On the basis of these facts, we consider that the smaller $c/a$ ratio and/or lattice strain toward a lower-symmetry structure such as monoclinic are essential for the appearance of higher-$T_c$ superconductivity in the $LaO_{0.5}F_{0.5}BiS_2$ superconductor. As an important fact, the high-$T_c$ phase of the $LaO_{0.5}F_{0.5}BiS_2$ superconductor can be obtained under the environment of high pressure and low reacting temperature. To analyze the local structure in detail, growth of the single crystal of the high-$T_c$ phase is required.

In conclusion, high-quality polycrystalline samples of $LaO_{0.5}F_{0.5}BiS_2$ were obtained using high-pressure synthesis technique. The $LaO_{0.5}F_{0.5}BiS_2$ sample prepared by heating at 700 °C for 1 h under 2 GPa showed the highest $T_c$ and shielding volume fraction. The estimated transition temperatures were $T_c^{onset}$ = 11.1 and $T_c^{zero}$ = 8.5 K in the electrical resistivity measurements and $T_c^{onset}$ = 11.5 and $T_c^{irr}$ = 9.4 K in the magnetic susceptibility measurements. These values are higher than those of the $LaO_{0.5}F_{0.5}BiS_2$ polycrystalline samples obtained from a conventional solid-state reaction. It was found that the high-$T_c$ phase can be stabilized under high pressure and relatively-low heating temperature. XRD experiments indicated that to realize the high-$T_c$ phase in $BiS_2$-based superconductor, the smallest value of the $c/a$ ratio of the lattice constants is essential. Since the superconducting properties of the HP-synthesized $LaO_{0.5}F_{0.5}BiS_2$ sample is better than those of the solid-state-reacted samples and the HP-post-annealed samples, detailed investigations of superconducting properties using the HP sample will provide us with important information for elucidating the mechanisms of $BiS_2$-based superconductivity.

Acknowledgements

The authors thank Dr. Y. Takano of National Institute for Materials Science and Dr. R. Higashinaka of Tokyo Metropolitan University for their fruitful comments on HP experiments. This work was partly supported by a Grant-in-Aid for Scientific Research for Young Scientists (A).

Table 1. Conditions of HP synthesis (annealing temperature and time) and superconducting transition temperatures of $T_c^{irr}$ and $T_c^{onset}$ examined from magnetic susceptibility measurements for $LaO_{0.5}F_{0.5}BiS_2$ polycrystalline samples.

| Synthesis conditions | | | $T_c^{irr}$ (K) | $T_c^{onset}$ (K) | Impurity |
|---|---|---|---|---|---|
| Solid-state reaction at 700 ºC | | | 2.5 | 3.5 | Single phase |
| HP (2 GPa) | 600 ºC | 0.5 h | 6.6 | 10.8 | Mixed phase |
| HP (2 GPa) | 650 ºC | 1.5 h | 9.1 | 11.1 | Mixed phase |
| HP (2 GPa) | 700 ºC | 0.5 h | 8.5 | 11.2 | Mixed phase |
| HP (2 GPa) | 700 ºC | 1 h | 9.4 | 11.5 | Single phase |
| HP (2 GPa) | 700 ºC | 2 h | 7.0 | 9.1 | Single phase |
| HP (2 GPa) | 800 ºC | 0.5 h | 4.1 | 8.6 | Single phase |

Figure captions

Fig. 1. (a) Temperature dependence of electrical resistivity for solid-state-reacted $LaO_{0.5}F_{0.5}BiS_2$ and HP-700 ºC $LaO_{0.5}F_{0.5}BiS_2$ (2 GPa, 700 ºC, 1 h). (b) Temperature dependence of magnetic susceptibility for solid-state-reacted $LaO_{0.5}F_{0.5}BiS_2$ and HP-700 ºC $LaO_{0.5}F_{0.5}BiS_2$. (c), (d) Enlargements of (a) and (b) around the onset and bifurcation temperatures of superconducting transition.

Fig. 2. Powder X-ray diffraction pattern of HP-700 ºC $LaO_{0.5}F_{0.5}BiS_2$.

Fig. 3(a) Enlargement of Fig. 2 around the (003) peaks with the X-ray diffraction results of solid-state-reacted $LaO_{0.5}F_{0.5}BiS_2$ and HP-800 ºC $LaO_{0.5}F_{0.5}BiS_2$. (b) Enlargement of Fig. 2 around the (200) peak with the X-ray diffraction results of solid-state-reacted $LaO_{0.5}F_{0.5}BiS_2$ and HP-800 ºC $LaO_{0.5}F_{0.5}BiS_2$..

Fig. 4. $T_c^{onset}$ and $T_c^{irr}$ plotted as a function of the *c/a* ratio. The data are from the results of the solid-state-reacted sample, HP-700 ºC sample and HP-800 ºC sample of $LaO_{0.5}F_{0.5}BiS_2$.

Figures

Fig. 1

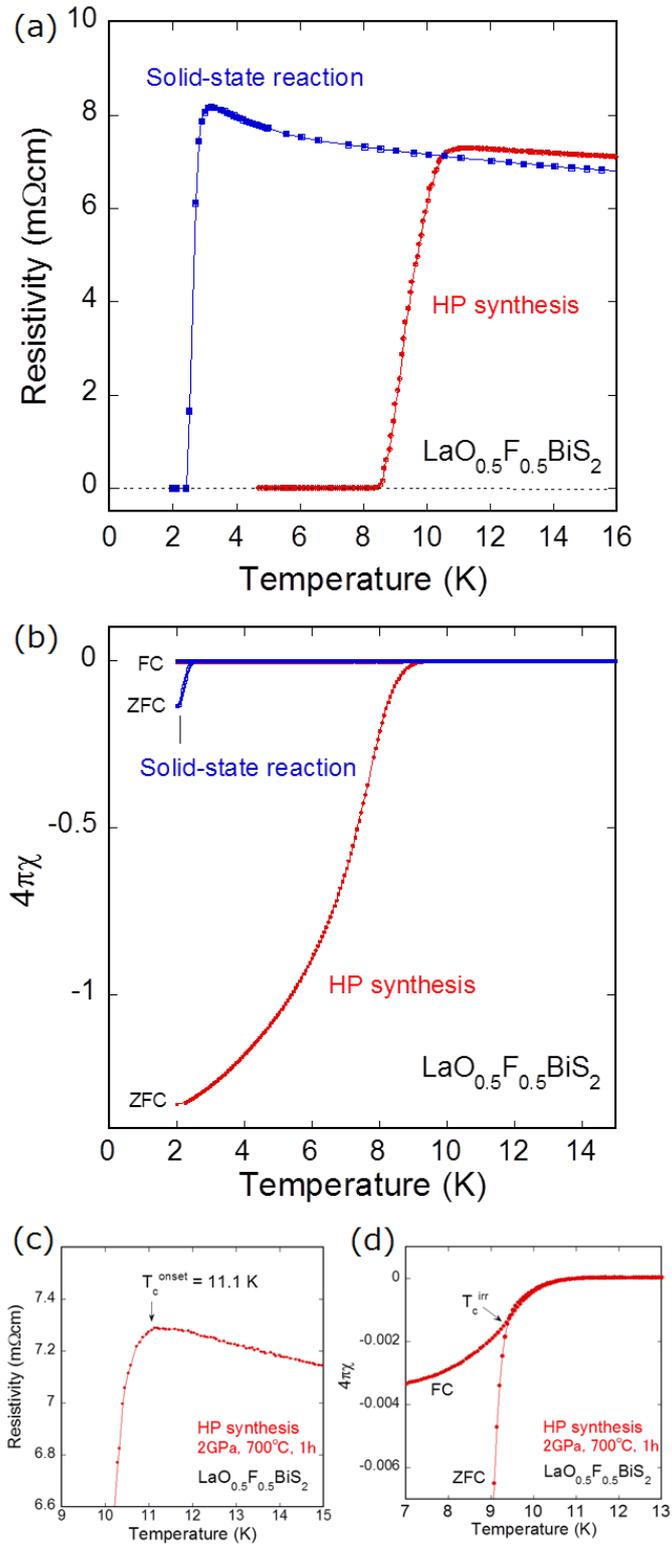

Fig. 2.

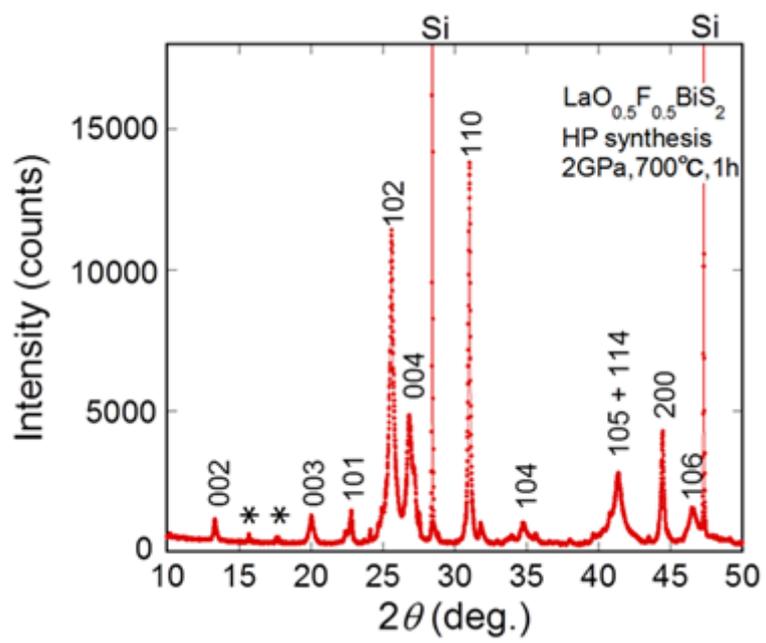

Fig. 3.

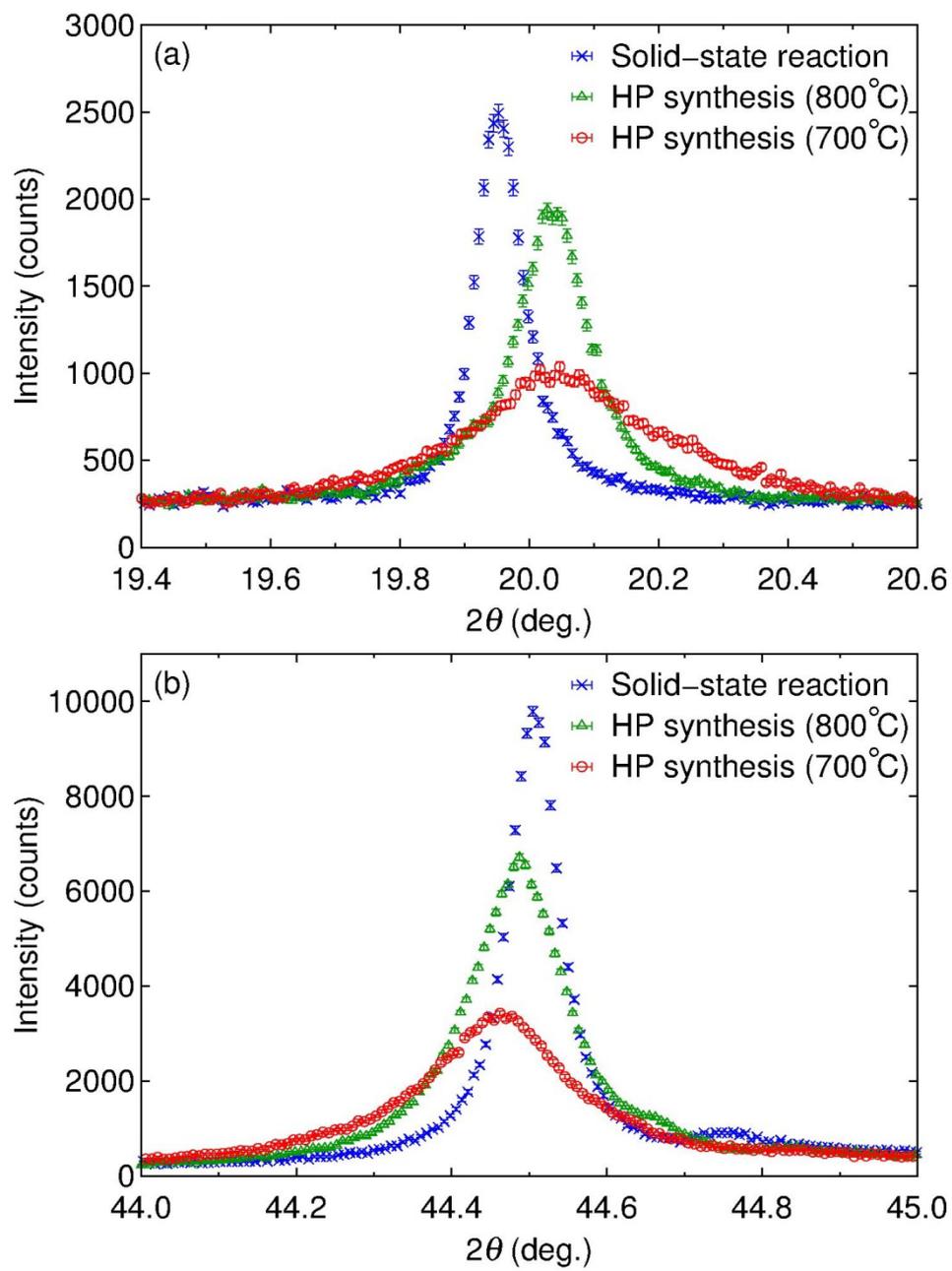

Fig. 4.

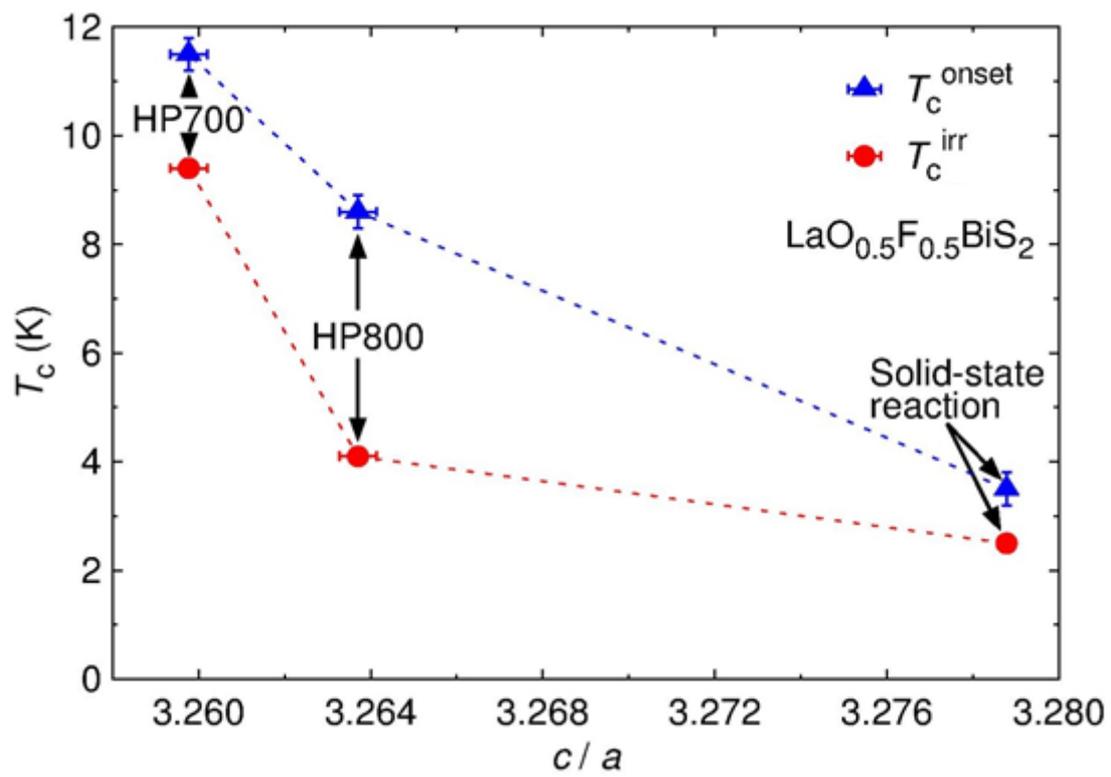